\documentstyle[aps,prb,multicol,epsf,picinpar]{revtex} 

\def\ep {\epsilon}

\def\e2 {\epsilon-\epsilon_k}
\def\be {\begin{equation}}
\def\ee {\end{equation}}
\def\bea {\begin{eqnarray}}
\def\eea {\end{eqnarray}}
\def\om {\omega}

\begin{document}
\draft           
\draft
\title{Quasiparticle scattering rate in overdoped superconducting cuprates}

\author{ George Kastrinakis}

\address{Institute of Electronic Structure and Laser (IESL), Foundation for
Research and Technology - Hellas (FORTH), 
P.O. Box 1527, Iraklio, Crete 71110, Greece$^*$}

\author{ Published in Phys. Rev. B {\bf 71}, 014520 (2005) }

\maketitle

\begin{abstract} 

We calculate the quasiparticle scattering rate in the superconducting
state of the overdoped cuprates, in the context of the Eliashberg formalism 
for a Fermi liquid with strong van Hove singularities close to the chemical 
potential. For a $d_{x^2-y^2}$ superconducting gap, we demonstrate 
analytically that the scattering
rate is linear in the maximum of temperature or energy, but with 
different intercepts and momentum dependence, thus extending our earlier
results on the normal state. 
We discuss our results in view of angle-resolved photoemission experiments. 
We also comment on the case of a s-wave gap.

\end{abstract}

\vspace{.3cm}

The nature of the carriers in the cuprates is an important question,
which has been discussed and debated extensively. The availability
of relevant experimental data over the years, and 
in particular angle-resolved photoemission (ARPES), 
should be helpful in the effort towards achieving a better
understanding of this question.
	
Here, we look at a characteristic one-particle property, relevant to ARPES.
We obtain analytically the scattering rate in the superconducting
state for a Fermi liquid with strong density of states peaks - van-Hove
singularities (vHs) in 2-D - located close to the chemical potential $\mu$.
Incidentally, this model
does not apply to the underdoped cuprates, as there is no indication
that a Fermi liquid description is appropriate. The {\em qualitative} nature
of our results does not depend on the doping level, for as long as we
remain within the realm of Fermi liquid theory. Precise numerical
calculations can yield the quantitative dependence of the theory on the
doping etc.
A review of related work in the frame of the so-called van-Hove 
scenario has been given in \cite{mark}. 
The pinning of the vHs close to $\mu$ seems to be a plausible explanation
for the common characteristics of a good many cuprates, whose van-Hove
singularities are located between 10-30 $meV$ {\em below} the Fermi surface
\cite{bednorz}, as ARPES experiments have shown.
A review of some calculations yielding the pinning of the vHs close to $\mu$ 
appears in \cite{gk2}. The present work extends our results for the normal
state, obtained in \cite{gk2,gk}, to the superconducting state. The analysis
presented below can also be done in the frame of a weak-coupling BCS-type 
approach. We opt for the intermediate to strong-coupling Eliashberg 
approach, which is relevant for the cuprates. {\em Qualitatively}, i.e. as
far as power laws etc. are concerned in terms of energy and temperature,
the answers are the same for the two approaches.

	We consider a generic dispersion appropriate for the cuprates,
of the type $t,t',t''$ 
\be
\epsilon_k = -2 t (\cos{k_x}+\cos{k_y})-4t'\cos{k_x}\cos{k_y}-
2t''(\cos{2k_x}+\cos{2k_y}) \;\; .
\ee
This dispersion generates strong vHs located at the points 
$(\pm\pi,0),(0,\pm\pi)$.

In the context of the Eliashberg formalism, 
the Green's function in the superconducting state is
\be
G(k,\ep_n)=\frac{i \ep_n Z(k,\ep_n)+\ep_k -\mu + X(k,\ep_n)}{P(k,\ep_n)}
\;\;,
\ee
with $\ep_n = (2 n +1)\pi T$ and
\be
P(k,\ep_n) = [i \ep_n Z(k,\ep_n)]^2-[\ep_k -\mu + X(k,\ep_n)]^2-D^2(k,\ep_n)
\;\;.
\ee
The renormalization $Z(k,\ep_n)$, the energy shift $X(k,\ep_n)$ and 
the (singlet) superconducting gap parameter $D(k,\ep_n)$ are given 
by the Eliashberg equations
\be
i \ep_n [1-Z(k,\ep_n)] = T \sum_{p,\ep_{n'}} \frac{ V(k-p,\ep_n-\ep_{n'}) 
i \ep_{n'} Z(p,\ep_{n'}) }{P(p,\ep_{n'})}   \;\;,
\ee
\be
X(k,\ep_n) = T \sum_{p,\ep_{n'}} \frac{ V(k-p,\ep_n-\ep_{n'}) \;
[\ep_p -\mu + X(p,\ep_{n'})] }{P(p,\ep_{n'})}   \;\; ,
\ee
and
\be
D(k,\ep_n) = T \sum_{p,\ep_{n'}} \frac{ V_p(k-p,\ep_n-\ep_{n'})
D(p,\ep_{n'})} {P(p,\ep_{n'})}   \;\;.
\ee
$V(q,\om_n)$ generates the diagonal self-energy - c.f.
below - and $V_p(q,\om_n)$ is the pairing potential. These potentials 
depend on the quasiparticle Green's function through screening. In 
general, e.g. for spin dependent interactions etc., these two potentials 
differ.

The diagonal self-energy is given by
\be
\Sigma(k,\ep_n) = i \ep_n [1-Z(k,\ep_n)] + X(k,\ep_n) 
= T \sum_{p,\ep_{n'}} V(k-p,\ep_n-\ep_{n'}) \; G(p,\ep_{n'}) \;\;.
\ee

Our derivation relies on this last equation, which is valid quite generally
in the frame of a Baym-Kadanoff (BK) conserving approximation \cite{bk},
irrespectively of the specific Hamiltonian and approximation
thereof \cite{gk2}. 

The quasiparticle scattering rate is twice 
the imaginary part of $\Sigma(k,\ep)=\Sigma_1(k,\ep)+i\Sigma_2(k,\ep)$. 
It can easily be shown \cite{agd} that $Im \Sigma(k,\ep)$
is given by the following formula for real energy at finite temperature :
\be 
\Sigma^R_2(k,\ep) = \sum_{q} \int_{-\infty}^{\infty} \frac{d\om}{2 \pi}
Im G^R(k-q,\ep-\om) Im V^R(q,\om) \; 
\{\coth(\om/2T) \; + \; \tanh((\ep-\om)/2T) \}\;\;.
\ee

We have
\be
Im G^R(k,\ep) = B(k,\ep) \; Im\Big\{ \frac{1}{a-b-D} \Big\} +
F(k,\ep) \; Re \Big\{ \frac{1}{a-b-D} \Big\} \;\;.
\label{fige}
\ee
Here 
$a=\ep Z(k,\ep), b=\ep_k-\mu+X(k,\ep), D=D(k,\ep)$, 
$B(k,\ep)=Re\Big\{ \Big(\frac{a+b}{a+b+D}\Big)^R \Big\}$
and $F(k,\ep)=Im \Big\{ \Big(\frac{a+b}{a+b+D}\Big)^R \Big\}$. Note that 
$a-b=G^{-1}(k,\ep)$ in the normal state.

We see that the factor
\be
F(k,\ep) = \frac{f_2(k,\ep) D_1(k,\ep) -f_1(k,\ep) D_2(k,\ep)}
{[f_1(k,\ep)+ D_1(k,\ep)]^2+[f_2(k,\ep)+ D_2(k,\ep)]^2} \;\;, 
\ee
with $f_1(k,\ep) = Re f(k,\ep)$, $f_2(k,\ep) = Im f(k,\ep)$,
$f(k,\ep)=\ep Z(k,\ep) +\ep_k-\mu+X(k,\ep)$, and
$D(k,\ep)=D_1(k,\ep)+iD_2(k,\ep)$.
This term is proportional to the gap function, and hence it gives
a negligibly small contribution upon summation over the Brillouin zone
for a $d_{x^2-y^2}$ gap $D_i(k,\ep) \simeq D_{io}(\ep) d(k)$,
$d(k)=\cos k_x - \cos k_y$. 
So we can ommit the second term in the r.h.s. of eq. (\ref{fige}). 
This is not the case for a $s$-wave gap though, where this term should 
be kept.

$B(k,\ep)$ is given by
\be
B(k,\ep) = \frac{f_1^2(k,\ep)+f_2^2(k,\ep)+ D_1(k,\ep)f_1(k,\ep)
+ D_2(k,\ep)f_2(k,\ep)}
{[f_1(k,\ep)+ D_1(k,\ep)]^2+[f_2(k,\ep)+ D_2(k,\ep)]^2} \;\;.
\ee

We can make the approximation
\be
Im \Big\{ \frac{1}{a-b-D} \Big\} \simeq -\pi
\delta(\ep+\mu-\ep_k- \Sigma_1(k,\ep) -  D_1(k,\ep)) \;\; .\label{gds}
\ee
This approximation was made in \cite{gk,gk2}.
Setting $Im G^R(k,\ep)$ equal to a delta function is a reasonable
approximation for this purpose, in view of the typical sharp spike feature
of $Im G^R(k,\ep)$ shown in \cite{gk2} - c.f. fig. 1 therein. 
This is in the frame of numerical
self-consistent FLEX calculations \cite{bic}.
Further, numerically $Im G^R(k,\ep)$ is 
{\em very small} compared to the band energy for small couplings,
and the difference of $Im \Sigma(k,\ep)$, as seen in our numerical
calculation, for small and large coupling constants is mostly 
{\em quantitative} rather than qualitative. 
Moreover, the approximation of eq. (\ref{gds}) has also been used in the 
context of the electron-phonon problem in \cite{agd} (c.f. the equation
right before eq. (21.29) therein).

Further, omitting terms of order
$[\om^2 \; \partial^2_{\ep} Re\Sigma(k,\ep)]$
and $[\om^2 \; \partial^2_{\ep} Re D(k,\ep)]$ we obtain
\be
\Sigma^R_2(k,\ep) \simeq - \frac{1}{2} \sum_{q} B(k-q,\ep-w_{k-q,\ep}) 
Im V^R(q,w_{k-q,\ep}) \; 
\{\coth(w_{k-q,\ep}/2T) \; + \; \tanh((\ep-w_{k-q,\ep})/2T) \}\;\;,
\ee
where
\be
w_{k,\ep} =\frac{\ep+\mu-\ep_k- \Sigma_1(k,\ep) - D_1(k,\ep)}
{1 - \partial_{\ep} \Sigma_1(k,\ep) - \partial_{\ep} D_1(k,\ep)}
\;\;.
\ee

	We write
\be
Im V^R(q,x) = \sum_{n=0}^{\infty} \; \frac{V_q^{(2n+1)}(0) \; x^{2n+1}}
{(2n+1)!} \; , \label{vexp}
\ee
where $V_q^{(n)}(0)$ is the $n-$th derivative of $Im V^R(q,\om=0)$ with respect
to $\om$. Eq. (\ref{vexp}) is true in the context of a BK approximation,
and $V(q,\om)$ may contain vertex corrections.
E.g. for RPA and FLEX we see that the overall behavior of $Im V$ closely
follows the carrier susceptibility(-ies) $\chi_o(q,\om)$, which 
we assume to be a regular function of $\om$, as
\be
Im V(q,\om) = \sum_i Im \chi_{oi}(q,\om) \;\; 
|\varepsilon_i(q,\om)|^2 \;\;. \label{imv}
\ee
$Im \chi_o$ is odd in $\om$, while $|\varepsilon|^2$ is even, 
with $\varepsilon(q,\om)$ the dielectric function.
In the FLEX approximation, $i=$spin,charge, in RPA $i=$charge.
In other BK approximations, e.g. c.f. \cite{tak}, eq. (\ref{vexp})
is valid as well, with $Im V(q,\om)$ obeying an equation similar 
to eq. (\ref{imv}).

The derivation given below for the
linear in energy and temperature behavior of $\Sigma_2(k,\ep;T)$
is generic. It relies essentially on a large coefficient for the
linear in energy term of $Im V$ - i.e. of $Im \chi_o$ etc. - and
the presence of van-Hove singularities
{\em near} the Fermi surface. The result holds {\em regardless} of the
dimensionality of the system. However, it is important that a significant
part of the spectral weight be included in the strong peaks of the density
of states lying close to $\mu$, a situation which is particularly favored
in 2-D.
 
For {\em sufficiently small} $V_{q}^{(n)}(0), \; \forall n>1$, we obtain
\be
\sum_{q } V_{k-q}^{(1)}(0) \;w_{k-q,\ep}\; \gg \; \sum_{q } 
\sum_{n=1}^{\infty} \; \frac{ V_{k-q}^{(2n+1)}(0) \; 
(w_{k-q,\ep})^{2n+1}}{(2n+1)!}  \label{eqv}
\;\;.
\ee
This relation is valid for $w_{k-q,\ep}< \ep_c$,
where the latter is the characteristic energy beyond which the infinite sum
on the right becomes comparable to the $V_{q}^{(1)}$ term
($\ep_c$ can be estimated by equating the lhs and rhs of eq.(\ref{eqv}) with
$n=1$ only).
We constrain ourselves to the regime where eq. (\ref{eqv}) is valid, as it
is precisely this regime which gives the linear in max\{$T,\ep$\}
dependence of $\Sigma_2$ in the normal state \cite{gk,gk2}.

Then we obtain
\be
\Sigma_2(k,\ep) \simeq -\frac{1}{2} 
\sum_q V_{q}^{(1)}(0) B(k-q,\ep-w_{k-q,\ep}) \; w_{k-q,\ep} \;
\{\coth(w_{k-q,\ep}/2T) \; + \; \tanh((\ep-w_{k-q,\ep})/2T) \}\;\;.
\ee

The sum over $q$ is dominated by the (extended) van Hove momenta 
$q_o$ in the vicinity
of $(\pm \pi,0)$ and $ (0,\pm \pi)$.

Now notice that the following inequalities are satisfied:
\bea
\ep+\mu-\ep_{k-q_o} > 0\;, \label{xe1} \\
Re \Sigma(k-q_o,\ep) < 0 \;\;.
\eea
The former is true for all momenta $k$ in the Brillouin zone 
except (possibly) for $k\simeq 0$.
The latter is seen to be true for e.g. $0\le\ep <4 t$ from numerical 
FLEX calculations - c.f. \cite{gk2,bic}. Combining the two we get
\be
\ep+\mu-\ep_{k-q_o} - Re \Sigma(k-q_o,\ep) > 0 \;\;.
\ee
Also, 
\be
\partial_{\ep} \Sigma_1(k,\ep) <0 \;\;,
\ee
and
\be
|\partial_{\ep} \Sigma_1(k,\ep)| \gg |\partial_{\ep} D_1(k,\ep)| 
\; \;, \label{xe2}
\ee
for $\delta < \ep\ <\ep_o$ \cite{gk2,bic}, where $\delta$ is a small
negative energy and $0< \ep_o = O(t)$. Overall eqs. (\ref{xe1})-(\ref{xe2})
yield
\be
w_{k-q_o,\ep} > 0 \;\;.
\ee

	We obtain a linear in $T$ scattering rate in the limit 
$T >\ep, w_{k-q,\ep} $
\be
\Sigma^R_2(k,\ep) \simeq -T \sum_{q } V_{q}^{(1)}(0) \;
B(k-q,\ep-w_{k-q,\ep}) 
\;\;. \label{tayt}
\ee
The coefficient of $T$ is maximized when $k-q_o$ lies along the diagonals
of the Brillouin zone, and minimized close to the vH points. This effect
is of order $(D_i/f_i)^2$ though, so it does not contribute much for a 
small gap.

	It has been known long ago - see e.g. \cite{sil}
- that the scattering rate becomes linear in $T$ for $T > \om_B/4$, with 
$\om_B$ being the characteristic boson frequency mediating the carrier 
interaction. Our treatment shows that $\om_B$ here is nothing else but
the {\em fermionic} energy $w_{k-q_o,\ep}$. The results here complement
our earlier normal state results in \cite{gk2,gk}.

ARPES experiments \cite{val,val2} 
have indicated that in the superconducting state, and, {\em presumably}, 
for $\ep \rightarrow 0$,
the scattering rate is linear in $T$ for $k_F$ close to the
nodal directions, but otherwise saturates to a value which increases as
$k_F$ approaches the vH points. Our result is in agreement with ARPES
for $k_F$ along the diagonals of the Brillouin zone. The reason for the 
off-diagonal discrepancy is not clear at present.
Certainly more data, and on compounds othen than BSCCO, would be helpful
in clarifying this issue.

We now turn to the limit $T < \ep$.
We note that for $\ep<w_{k-q,\ep} $ and for $w_{k-q,\ep}<0 $ the contributions
of tanh and coth annihilate each other in the low $T$ limit. 
Then, for $\ep>w_{k-q_o,\ep} $
\be
\Sigma^R_2(k,\ep) \simeq - \frac{c_{k,\ep}}{2} \sum_{q } V_{q}^{(1)}(0) \;
B(k-q,\ep-w_{k-q,\ep}) \; w_{k-q,\ep}    \label{exen}
\;\;,
\ee
where $c_{k,\ep} =1+ <\tanh((\ep-w_{k-q_o,\ep})/2T)>$ is a factor
between 1 and 2.
This equation yields a linear in $\ep$ scattering
rate, if $\partial_{\ep} \Sigma_1(k-q_o,\ep)$ 
varies slowly with $\ep$. Actually, assuming that 
$D_i^2(k,\ep)\ll f_i^2(k,\ep)$ we obtain
\be
\Sigma^R_2(k,\ep) \simeq - \frac{2 V_{q_o}^{(1)}(0) N_o c_{k,\ep}}
{1-\partial_{\ep} \Sigma_1(k',\ep)} \Big\{ 
\ep+\mu-\ep_{k'}-\Sigma_1(k',\ep) - 
\frac{[f_1(k',w) D_{1o}(w)+f_2(k',w) D_{2o}(w)]D_{1o}(\ep) d^2(k')}
{f_1^2(k',w)+f_2^2(k',w)}  \Big\}  \label{exse}
\;\;.
\ee
where $k'=k-q_o$, $w=\ep-w_{k-q_o,\ep}$ and $N_o$ stands for the q-summation
around a single vHs.
We see that this expression is linear in $\ep$, and that the gap-dependent
term is minimized for $k-q_o$ lying along the diagonals (with the assumption
that $f_i(k',w)$ vary more slowly with $k'$ than the gap).
For a s-wave gap we obtain a dependence $\ep - D$ instead, just from the
term $w_{k-q_o,\ep}$ in eq. (\ref{exen}), 
and $B(k,\ep)$ yields an additional $D$ contribution.
These two distinct dependencies of $\Sigma_2$ on the gap are characteristic
of the respective gap symmetries.
Estimates of the parameters in eq. (\ref{exse}) cannot be easily obtained
analytically, due to the complicated nature of the Eliashberg formalism.
Precise numerical estimates are possible, but they are beyond the scope 
of the present work.

ARPES experiments in \cite{jo1,jo2,val2,arg} indicate that the scattering rate
is a linear function of energy both in the superconducting and 
the normal state, in agreement with our results. The normal state result
was obtained in \cite{gk2,gk}.

Summarizing, within the Fermi liquid Eliashberg formalism, we calculate
the one-particle scattering rate in the superconducting state of 
overdoped cuprates. We compare our findings with existing ARPES 
experiments. For $T<\ep$, the $\ep$-dependence obtained
is consistent with expts. For $T>\ep$, the $T$-dependence obtained 
agrees with expts. for $k_F$ along the Brillouin zone diagonals.

\vspace{.2cm}
The author acknowledges useful correspondence with M.R. Norman on ARPES data.

\vspace{1cm}
$^*$ E-mail: kast@iesl.forth.gr .


\begin{references}

\bibitem{mark}
R.S. Markiewicz, J. Phys. Chem. Sol. {\bf 58}, 1179 (1997); cond-mat/9611238.

\bibitem{bednorz}
D.H. Lu et al., Phys. Rev. Lett. {\bf 76}, 4845 (1996).

\bibitem{gk2}
G. Kastrinakis, Physica C {\bf 340}, 119 (2000); Erratum, ibid {\bf 382},
445 (2002).

\bibitem{gk}
G. Kastrinakis, 
Physica C {\bf 317-319}, 497 (1999).

\bibitem{bk}
G. Baym, Phys. Rev. {\bf 127}, 1391 (1962); G. Baym and L.P. Kadanoff, ibid.
{\bf 124}, 287 (1961).

\bibitem{agd}
A.A. Abrikosov, L.P. Gorkov and I.E. Dzyaloshinski, {\em Methods of Quantum 
Field Theory in Statistical Physics}, Prentice-Hall (1964).

\bibitem{bic}
C.H. Pao and N.E. Bickers, Phys. Rev. B {\bf 51}, 16310 (1995).

\bibitem{tak}
Y. Takada, Phys. Rev. B {\bf 52}, 12708 (1995).

\bibitem{sil}
Q. Si \& K. Levin, Phys. Rev. B {\bf 44}, 4727 (1991).

\bibitem{val}
T. Valla, A.V. Fedorov, P.D. Johnson, Q. Li, G. D. Gu and N. Koshizuka,
Phys. Rev. Lett. {\bf 85}, 828 (2000).

\bibitem{val2}
T. Valla, A.V. Fedorov, P.D. Johnson, B.O. Wells, S.L. Hulbert, 
Q. Li, G.D. Gu, and N. Koshizuka, Science {\bf 285}, 2110 (1995).

\bibitem{jo1}
Z.M. Yusof, B.O. Wells, T. Valla, A.V. Fedorov, P.D. Johnson, Q. Li, 
C. Kendziora, S. Jian, and D. G. Hinks, 
Phys. Rev. Lett. {\bf 88}, 167006 (2002).

\bibitem{jo2}
P.D. Johnson, T. Valla, A.V. Fedorov, Z. Yusof, B.O. Wells, Q. Li, 
A.R. Moodenbaugh, G.D. Gu, N. Koshizuka, C. Kendziora, S. Jian 
and D. G. Hink, Phys. Rev. Lett. {\bf 87}, 177007 (2001).

\bibitem{arg}
A. Kaminski, M. Randeria, J.C. Campuzano, M.R. Norman, H. Fretwell, 
J. Mesot, T. Sato, T. Takahashi, and K. Kadowaki, Phys. Rev. Lett. {\bf 86},
1070 (2001).

\end{references}
\end{document}